\documentclass[%
 reprint,
 amsmath,amssymb,
 aps,
]{revtex4-2}

\usepackage{graphicx}
\usepackage{dcolumn}
\usepackage{bm}

\begin{document}

\title{Optical Properties of Charged Defects in Monolayer MoS$_2$}
\author{Martik Aghajanian}
\author{Arash A. Mostofi}
\author{Johannes Lischner}
\email{j.lischner@imperial.ac.uk}
\affiliation{Departments of Physics and Materials and the Thomas Young Centre for Theory and Simulation of Materials, Imperial College London, London, SW7 2AZ, UK}

\date{\today}
\begin{abstract}
We present theoretical calculations of the optical spectrum of monolayer MoS$_2$ with a charged defect. In particular, we solve the Bethe-Salpeter equation based on an atomistic tight-binding model of the MoS$_2$ electronic structure which allows calculations for large supercells. The defect is modelled as a point charge whose potential is screened by the MoS$_2$ electrons. We find that the defect gives rise to new peaks in the optical spectrum approximately 100-200 meV below the first free exciton peak. These peaks arise from transitions involving in-gap bound states induced by the charged defect. Our findings are in good agreement with experimental measurements.
\end{abstract}
\maketitle

\section{Introduction}
Monolayer transition-metal dichalcogenides (TMDs) are two-dimensional (2D) materials which have been intensely studied in recent years because of their attractive electronic properties for applications in transport and optoelectronic devices~\cite{Wachter2017, LopezSanchez2013, Baugher2014}. Many materials in this class exhibit a direct band gap in the optical range as well as multiple band extrema~\cite{liu15, kuc15, Mak2010} which give rise to rich valley physics~\cite{Cao2012}. The reduced dimensionality of 2D materials causes a weaker electronic screening of electron-electron interactions compared to 3D systems and the stronger interaction results in large binding energies of excitons, bound electron-hole pairs. In monolayer TMDs, exciton binding energies can be as large as several hundred meV~\cite{Colloquium2018}. 

Charged defects can have a significant impact on the electronic structure and transport properties of TMDs. In particular, doped carriers can increase the conductivity, while scattering from charged defects reduces it. The optical properties of TMDs are also influenced by the presence of charged defects. For example, Greben and coworkers~\cite{greben2020} demonstrated that irradiating monolayer MoS$_2$ with an electron beam gives rise to an additional peak (approximately 200 meV below the first neutral exciton peak) in the photoluminescene spectrum, which they interpreted as the signature of neutral excitons that are bound to an ionized donor defect. Similarly, Shang et al.~\cite{shang2017} studied the optical properties of monolayer WS$_2$ and MoS$_2$ and found that the photoluminescence could be tuned from donor-bound to acceptor-bound excitons by changing from n-doping to p-doping. 

To gain insight into the microscopic properties of excitons bound to charged defects, Ganchev and coworkers~\cite{ganchev2015} solved a three-particle Schroedinger equation based on an effective mass approximation for the electronic structure of monolayer TMDs. Similarly, Wu~\cite{wu2022} used the effective mass approximation to study transitions between bound defect states. However, such models do not capture the delicate effects associated with bound defect states arising from the multi-valley electronic structure of TMDs. To address this shortcoming of effective mass methods which typically only capture defect states from the $K$ and $K'$ valleys, we previously developed an atomistic approach to describe the electronic structure of a TMD monolayer with a charged defect~\cite{Aghajanian2018}. In particular, we used the tight-binding approach to model the large supercells required to describe the long-ranged electrostatic potential of the charged defect. Our calculations demonstrated that the most strongly bound acceptor states derive from the $\Gamma$ valley for a wide range of dielectric environments and defect charges. These predictions were verified by scanning tunneling spectroscopy experiments~\cite{Aghajanian2020}. For donor impurities, we predicted that the most strongly bound in-gap states derive from the $Q$ valleys for a range of dielectric environments and defect charges~\cite{Aghajanian2018}.

In this paper, we extend our atomistic modelling approach to calculate the optical properties of monolayer TMDs with charged defects. For this, we solve the Bethe-Salpeter equation using the tight-binding states as input. We calculate optical spectra of both donor and acceptor defects in MoS$_2$ on a SiO$_2$ substrate~\cite{Poellmann2015, Rigosi2016, Zhang2015}. We find that the charged defects induce additional low-energy peaks in the optical spectrum. These arise from electronic transitions which involve bound defect states. The binding energy of these excitations, which can be interpreted as defect-bound excitons, are between 100 and 200 meV in good agreement with experimental findings.

\section{Methods}

\subsection{Bethe-Salpeter equation}
To study the effect of a charged adsorbate on the optical properties of monolayer MoS$_2$, we solve the Bethe-Salpeter equation (BSE) for an $N\times N$ MoS$_2$ supercell with a single adsorbate, which is modelled as a point charge that creates a screened potential acting on the electrons in the MoS$_2$. 

The BSE is given by 
\begin{equation}
    \sum_{c'v'\mathbf{k'}} H^{\text{BSE}}_{cvc'v'}(\mathbf{k},\mathbf{k}') A^{M}_{c'v'\mathbf{k}'} = E_{M}A^{M}_{cv\mathbf{k}},
\end{equation}
where $E_M$ denotes the energy of the $M$-th excited state and $A^{M}_{cv\mathbf{k}}$ is the corresponding eigenvector. Here, $c$ and $v$ label conduction and valence states, respectively, and $\mathbf{k}$ is a crystal momentum in the first Brillouin zone.

The BSE Hamiltonian is given by \cite{Wu2015,Ridolfi2018}
\begin{align}
    \label{eqn:methods_bse_hamiltonian}
    H^{\text{BSE}}_{cvc'v'}(\mathbf{k},\mathbf{k}')  & = \delta_{vv'}\delta_{cc'}\delta_{\mathbf{kk'}}(E_{c\mathbf{k}}-E_{v\mathbf{k}}) \nonumber\\
    & - \left[ D^{cc'}_{vv'}(\mathbf{k},\mathbf{k'})-X^{cc'}_{vv'}(\mathbf{k},\mathbf{k'}) \right],
\end{align}
where $E_{n\mathbf{k}}$ denotes the energy of a quasiparticle state, with corresponding wavefunction $\psi_{n\mathbf{k}}(\mathbf{x})$ where $\mathbf{x}=(\mathbf{r},\alpha)$ comprises both a position $\mathbf{r}$ and a spin variable $\alpha$, and $D$ and $X$ are the direct and exchange integrals, respectively, given by
\begin{align}
&D^{cc'}_{vv'}(\mathbf{k},\mathbf{k'}) =
\nonumber \\ & \int\mathrm{d}\mathbf{x}\int \mathrm{d}\mathbf{x}'\psi_{v\mathbf{k}}(\mathbf{x})\psi^*_{c\mathbf{k}}(\mathbf{x'})W(\mathbf{r},\mathbf{r'})\psi^*_{v'\mathbf{k'}}(\mathbf{x})\psi_{c'\mathbf{k'}}(\mathbf{x'}), \\
& X^{cc'}_{vv'}(\mathbf{k},\mathbf{k'}) = \nonumber \\
&\int\mathrm{d}\mathbf{x} \int \mathrm{d}\mathbf{x}'\psi_{v\mathbf{k}}(\mathbf{x})\psi^*_{c\mathbf{k}}(\mathbf{x})v(\mathbf{r},\mathbf{r'})\psi^*_{v'\mathbf{k'}}(\mathbf{x'})\psi_{c'\mathbf{k'}}(\mathbf{x'}).
\end{align}
Here, $W(\mathbf{r},\mathbf{r}')$ and $v(\mathbf{r},\mathbf{r}')=e^2/(\varepsilon_\text{bg}|\mathbf{r}-\mathbf{r}'|)$ denote the screened and bare Coulomb interaction, respectively, with $\varepsilon_\text{bg}$ being the background dielectric constant and $e$ the proton charge. For a MoS$_2$ layer placed on a substrate material with dielectric constant $\varepsilon_\text{sub}$, we use $\varepsilon_\text{bg}=(\varepsilon_\text{sub}+1)/2$. The screened interaction in real space is obtained using a Hankel transform according to 
\begin{equation}
    \label{eq:hankel}
    W(r=|\mathbf{r}-\mathbf{r}'|) = e^2\int_0^{\infty}\mathrm{d}q\;\frac{e^{-qd}J_0(qr)}{\varepsilon_\text{bg} + \varepsilon_{\text{2D}}(q) },
\end{equation}
where $J_0(x)$ is the zeroth order Bessel function of the second kind and $\varepsilon_\text{2D}(q)$ is the 2D dielectric function of MoS$_2$, which is calculated from first-principles DFT with the random-phase approximation~\cite{Qiu2016,Aghajanian2018}. In the above, the parameter $d$ regularizes the divergence of the screened interaction when the electron and the hole reside on the same atom. We have found that $d=1.2$~\AA~ reproduces the experimentally measured binding energy of the lowest exciton~\cite{Wu2015}.

To efficiently calculate the quasiparticle energies and wavefunctions of MoS$_2$ with a charged adsorbate, we use the tight-binding (TB) approach. Following Liu and coworkers~\cite{liu13}, we express the wavefunctions as a linear combination of Mo $4d_{z^2}$, $4d_{xy}$ and $4d_{x^2-y^2}$ orbitals according to 
\begin{equation}
    \label{eq:basis}
    \psi_{n\mathbf{k}}(\mathbf{x}) = \frac{1}{\sqrt{N_k}} \sum_{lj}c_{n\mathbf{k}lj}
    \sum_{\mathbf{R}}e^{i\mathbf{k}\cdot(\mathbf{R} + \bm{\tau}_j)}\phi_l(\mathbf{r}-\mathbf{R}-\bm{\tau}_j,\alpha) ,
\end{equation}
where $\phi_l$ denotes an atomic basis function, $\mathbf{R}$ is a lattice vector and $\bm{\tau}_j$ denotes the position of the $j$-th atom relative to the origin of the supercell. Also, $N_k$ denotes the number of k-points used to sample the first Brillouin zone of the supercell and $c_{n\mathbf{k}lj}$ are complex coefficients obtained by diagonalizing the TB Hamiltonian. 

The TB Hamiltonian of MoS$_2$ with a charged adsorbate is constructed by starting from the Hamiltonian of pristine MoS$_2$ of Liu and coworkers, which has been fitted to reproduce the ab initio DFT band structure~\cite{liu13}, and create an $18\times 18$ supercell. Next, the screened potential induced by the charged adsorbate is added as an onsite potential. We assume that the defect has a charge of $Ze$ (with $Z=\pm1$) and is located at $x=y=0$ at a distance $D$ above a Mo atom. The corresponding screened potential is then given by $ZW(r)$ with $d$ in Eq.~\eqref{eq:hankel} replaced by $D$.

Inserting the TB ansatz for the quasiparticle wavefunctions into the exchange and direct integrals and exploiting the localization of the atomic basis functions yields
\begin{align}
D^{cc'}_{vv'}(\mathbf{k},\mathbf{k}') & = \sum_{ij}\left(T^{(i)}_{c\mathbf{k},c'\mathbf{k'}}\right)^* W_{ij}(\mathbf{k}-\mathbf{k'})T^{(j)}_{v\mathbf{k},v'\mathbf{k'}} ,\\
X^{cc'}_{vv'}(\mathbf{k},\mathbf{k'})  &= \sum_{ij}\left(T^{(i)}_{c\mathbf{k},v\mathbf{k}}\right)^* v_{ij}(0)T^{(j)}_{c'\mathbf{k'},v'\mathbf{k'}},
\end{align}
where we define $T^{(j)}_{n\mathbf{k},n'\mathbf{k'}} =\sum_l c_{n\mathbf{k}lj}c^*_{n'\mathbf{k'}lj}$,  $W_{ij}(\mathbf{k})=\sum_{\mathbf{R}}\exp(-i\mathbf{k}\cdot\mathbf{R})W(\mathbf{R}+\bm{\tau}_i - \bm{\tau}_j)$ and $v_{ij}(\mathbf{k})=\sum_{\mathbf{R}}\exp(-i\mathbf{k}\cdot\mathbf{R})v(\mathbf{R}+\bm{\tau}_i - \bm{\tau}_j)$. We have found that the effect of exchange interactions on the absorption spectrum is small (see Fig. 4 in the Appendix) and have therefore neglected it in our calculations. 

As the size of the BSE Hamiltonian increases rapidly with the supercell size, we only include those conduction states at each k-point which fulfill $E_{c\mathbf{k}} \leq E_{v\mathbf{k}}^{\text{max}}+E_{\text{cut}}$ with $E_\text{cut}$ being a cutoff parameter and $E_{v\mathbf{k}}^{\text{max}}$ being the highest valence band energy at $\mathbf{k}$. Similarly, we only include valence states with $E_{v\mathbf{k}} \geq E_{c\mathbf{k}}^{\text{min}}-E_{\text{cut}}$ with $E_{c\mathbf{k}}^{\text{min}}$ denoting the lowest conduction band energy at $\mathbf{k}$. We then increase $E_\text{cut}$ until the energies of the lowest excitons are converged. The resulting cutoffs are shown in Table~\ref{tab:energy_cutoffs}. In our calculations, we use $\Gamma$ point sampling ($N_k=1$) of the first Brillouin zone associated with the supercell.

\begin{table}[]
    \centering
    \begin{tabular}{c|c|c|c}
        $Z$ & $\varepsilon_{\text{sub}}$ & $E_{\text{cut}}$ (eV) & $N_{\text{BSE}}$\\
        \hline
        -1 & 3.8 & 2.13 & 46,253 \\
        +1 & 3.8 & 2.08 & 46,011 \\
    \end{tabular}
    \caption{Cutoff energies $E_{\text{cut}}$ used for the BSE calculations. Note that the tight-binding band gap in our tight-binding model is $E_{\text{g}}=1.585$~eV. The linear dimension $N_\text{BSE}$ of the resulting BSE Hamiltonian is also given.}
    \label{tab:energy_cutoffs}
\end{table}

The real part of the optical conductivity is obtained from the eigenvectors and eigenvalues of the BSE according to~\cite{Ridolfi2018}
\begin{equation}
\label{eqn:methods_optical_cond_bse}
    \text{Re } \sigma_{xx}(\omega) =\frac{e^2}{\hbar m^2_0A}\sum_M \frac{\left|\sum_{\mathbf{k},c,v} A^{M}_{cv\mathbf{k}} \mathbf{\hat{x}}\cdot\mathbf{p}_{cv\mathbf{k}}\right|^2}{E_{M}} \delta(\hbar\omega-E_{M}),
\end{equation}
where $A=N_k A_\text{SC}$ (with $A_\text{SC}$ being the area of the supercell), $m_0$ denotes the bare electron mass and $\mathbf{e}=\mathbf{\hat{x}}$ is the polarization direction of the electric field of the electromagnetic wave with frequency $\omega$. The delta function is approximated by a normalized Lorentzian function with a full width at half maximum of $0.04$~eV. The momentum matrix elements are given by~\cite{Pedersen2001, Cruz1999, Ridolfi2018}
\begin{equation}
    \label{eq:momentum_matrix_elem}
    \mathbf{p}_{cv\mathbf{k}} = \frac{m_0}{\hbar}\sum_{limj}c^{*}_{c\mathbf{k}li}c_{v\mathbf{k}mj}\nabla_{\mathbf{k}}H^\text{TB}_{limj}(\mathbf{k}),
\end{equation}
where $H^\text{TB}_{limj}(\mathbf{k})$ denotes the tight-binding Hamiltonian in the atomic orbital basis, see Appendix for details. 

As we do not include a GW correction to our quasiparticle energies, it is necessary to shift the calculated optical spectrum such that the peak associated with the $A$ exciton agrees with the experimental value of 1.93~eV~\cite{Colloquium2018,Wu2015}. We have used this to align the spectrum both with and without the defect.

\section{Results}

\subsection{Quasiparticle states}

\begin{figure}
    \centering
    \includegraphics[width=0.49\textwidth]{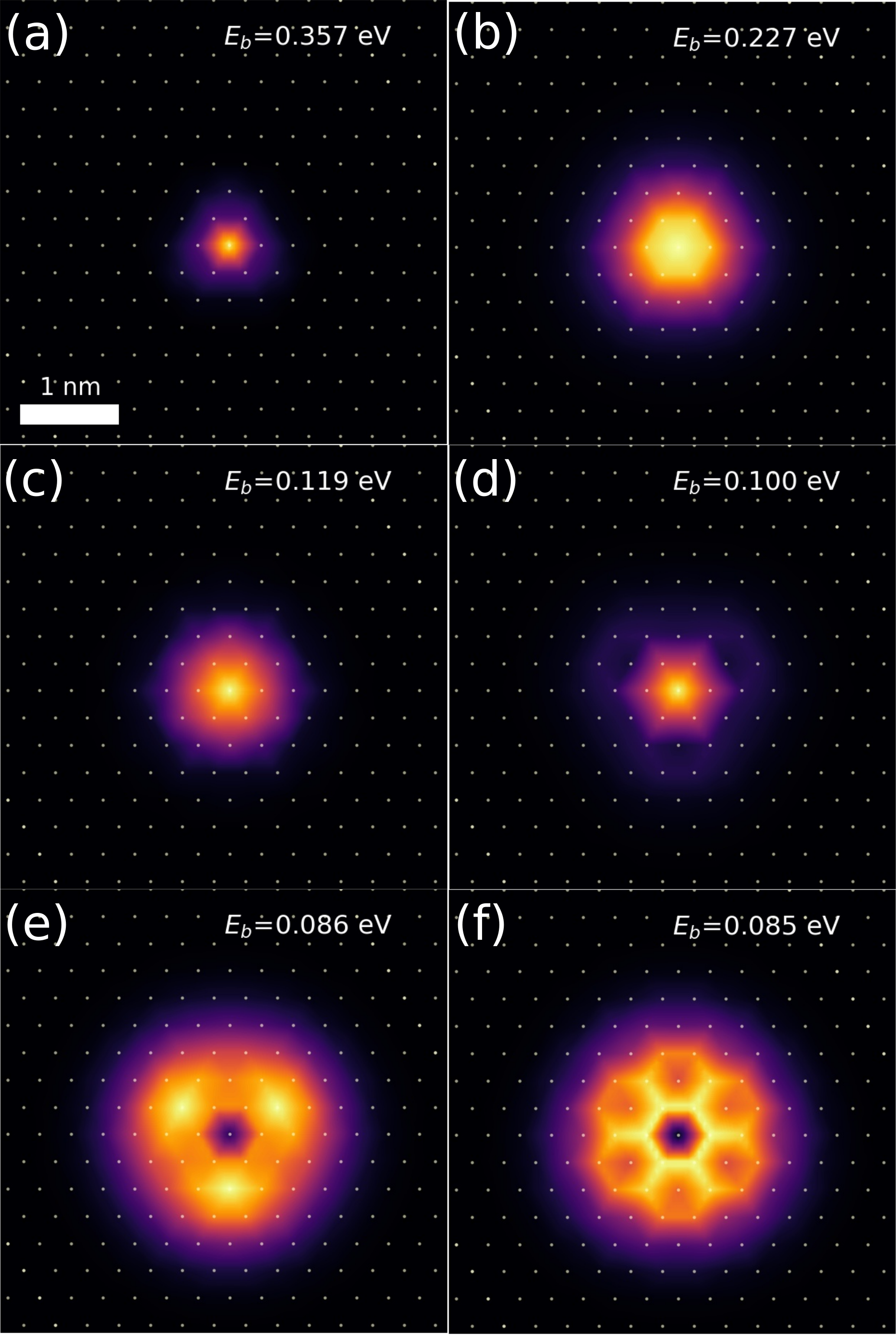}
    \caption{Squared magnitude of the wavefunctions of the most strongly bound acceptor states for a defect with charge $-|e|$ placed $2\;\text{\AA}$ above the Mo site of MoS$_2$ (on a SiO$_2$ substrate). Small dots indicate positions of Mo atoms.} 
    \label{fig:qp_states_acceptor}
\end{figure}

\begin{figure}
    \centering
    \includegraphics[width=0.49\textwidth]{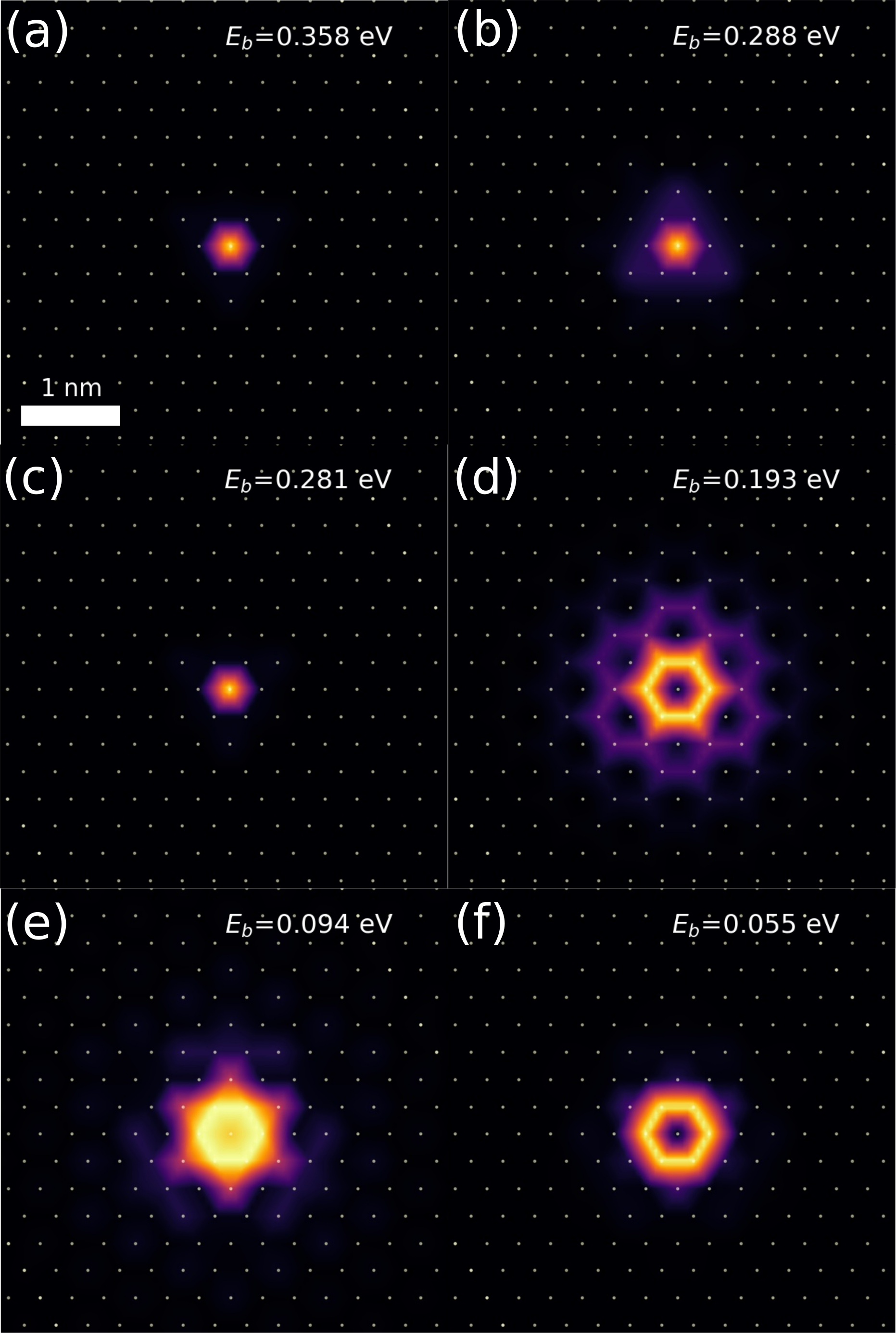}
    \caption{Squared magnitude of the wavefunctions of the most strongly bound donor states for a defect with charge $|e|$ placed $2\;\text{\AA}$ above the Mo site of MoS$_2$ (on a SiO$_2$ substrate). Small dots indicate positions of Mo atoms.}
    \label{fig:qp_states_donor}
\end{figure}

To understand the effect of charged adsorbates on the optical properties of MoS$_2$, we first discuss the quasiparticle states of this system. This discussion follows closely our previous work~\cite{Aghajanian2018}. Charged defects induce localized bound states in the band gap of the MoS$_2$. Fig.~\ref{fig:qp_states_acceptor}(a-f) shows the real-space wavefunctions of the most strongly bound acceptor states induced by a negatively charged adsorbate. We have used $\varepsilon_{\text{sub}}=3.8$ corresponding to a SiO$_2$ substrate. The most strongly bound defect state is highly localized and has 1s symmetry (Fig.~\ref{fig:qp_states_acceptor}(a)). It is composed of states from the $\Gamma$ valley of the MoS$_2$ band structure, even though the valence band maximum of the pristine material is located at the $K/K'$ points, as shown in Fig.~\ref{fig:acc_proj} in the Appendix. The $\Gamma$ valley has a large effective mass which gives rise to a highly localized state with a large binding energy. The next state also has 1s symmetry (Fig.~\ref{fig:qp_states_acceptor}(b), but is less localized. It is composed of states from the MoS$_2$ $K$ and $K'$ valleys which have a smaller effective mass than the $\Gamma$ valley. The state shown in Fig.~\ref{fig:qp_states_acceptor}(c) has a similar shape as the state in (b). Indeed, this state originates from the lower of the spin-split valence bands at K and K'. The other states in Fig.~\ref{fig:qp_states_acceptor} correspond to 2s and 2p states derived from the $\Gamma$ valley.

The wavefunctions of a donor defect are shown in Fig.~\ref{fig:qp_states_donor}. Again, we find that the three most strongly bound defect states are of 1s symmetry and highly localized. The most strongly bound donor states is composed of monolayer states from the $Q$ valleys, as shown in Fig.~\ref{fig:don_proj} of the appendix. Since the conduction band spin-orbit splitting is very small, the defect states from different $Q$ valleys of the Brillouin zone can hybridize and form different linear combinations whose energy splitting is determined by the Fourier component of the defect potential whose wave vector connects the different $Q$ valleys. In contrast, the state in Fig.~\ref{fig:qp_states_donor}(b) is a linear combination of 1s donor states from the $K$ and $K'$ valleys. The less strongly bound defect states, again, correspond to higher energy hydrogenic orbitals (and their linear combinations) from the $Q$ valleys and the $K/K'$ valleys.

\begin{figure*}[ht!]
    \centering
    \includegraphics[width=1.0\textwidth]{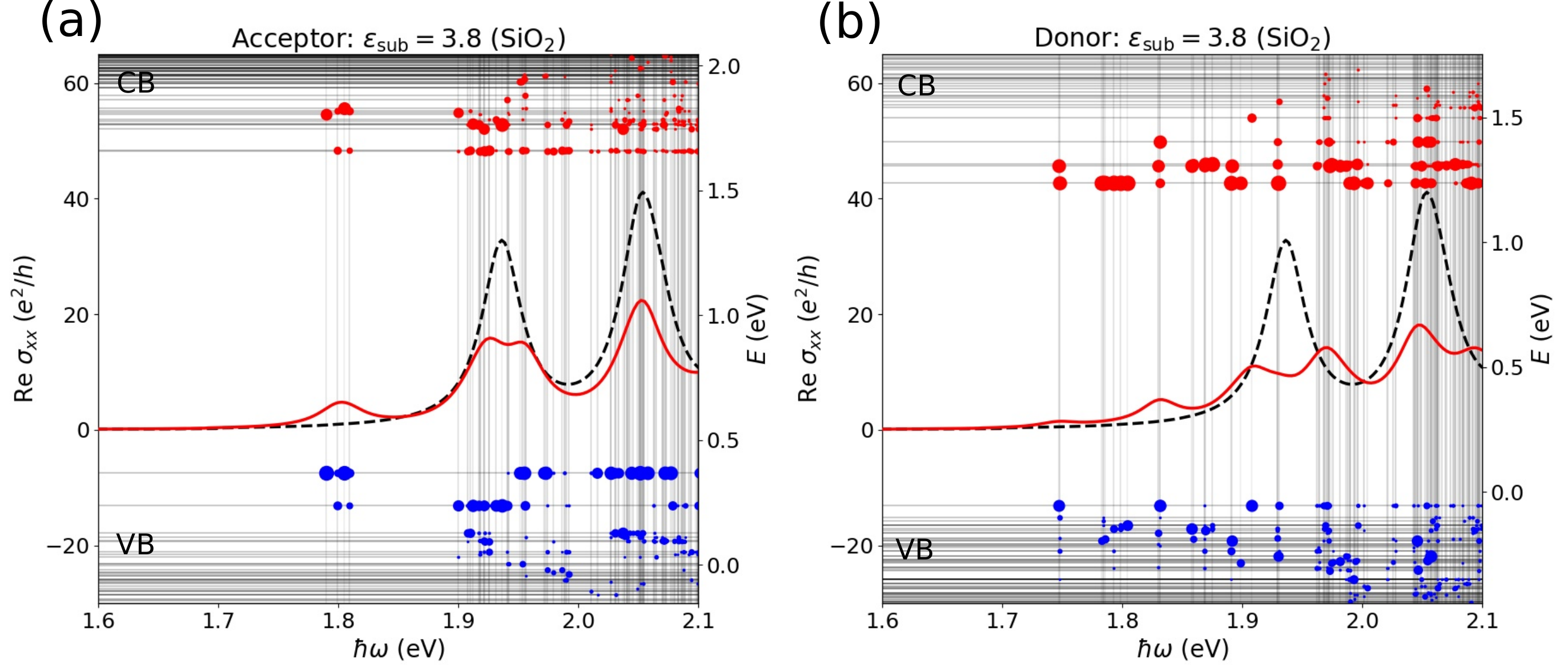}
    \caption{Optical conductivity of MoS$_2$ on a SiO$_2$ substrate with an acceptor (a) and donor (b) defect. Also shown are the projections of the excitonic states on the valence (blue dots) and conduction (red dots) quasiparticle states. The size of the dots is proportional to the squared magnitude of the projection of the exciton wavefunction onto the quasiparticle states.  
    } 
    \label{fig:opt_defect_cond_don}
\end{figure*}

\subsection{Optical properties}

The optical conductivity of monolayer MoS$_2$ on a SiO$_2$ substrate in the presence of an acceptor defect ($Z=-1$) is shown in Fig.~\ref{fig:opt_defect_cond_don}(a) and compared to result for the pristine defect-free material. Without defects, the conductivity is characterized by two large peaks at approximately $1.93$~eV and $2.05$~eV, corresponding to the well-known $A$ and $B$ excitons from the $K$ and $K'$ valleys. The energy difference between the two peaks reflects the spin splitting of the highest valence bands in these valleys from spin-orbit coupling. In the presence of the defect, the $A$ and $B$ exciton peaks are still present in the optical spectrum, but with significantly reduced intensities. Also, the $A$ peak is now split into two overlapping peaks. In addition, a new smaller peak arises at $\sim 1.8$~eV, i.e., at an energy approximately 130 meV lower than the $A$ exciton peak. 

To understand these findings, we also plot the squared magnitudes of the projections of the BSE eigenvectors $A^M_{cv}(\mathbf{k=0})$ onto the quasiparticle states, see Fig.~\ref{fig:opt_defect_cond_don}(a). This reveals that the new low-energy peak originates from several excitons which are predominantly composed of transitions from the most strongly bound defect state (of 1s character composed of $\Gamma$ valley states) to conduction band states. Transitions from the second most strongly bound defect state (of 1s character composed of $K$/$K'$ valley states) make a smaller contribution to the peak. Interestingly, the $A$ and $B$ peaks also contain transitions involving low-lying defect states.  

For MoS$_2$ on SiO$_2$ with a donor impurity, see Fig.~\ref{fig:opt_defect_cond_don}(b), the optical conductivity exhibits more peaks than for an acceptor impurity. In particular, both the $A$ and the $B$ peak of the pristine spectrum break into several smaller peaks. In contrast to the case of the acceptor impurity, we now observe two low-energy peaks: one at approximately $1.75$~eV and another one at approximately $1.83$~eV. The lowest peak is dominated by transitions from the valence band maximum to the two most strongly bound defects states. The peak at 1.83~eV involves transitions from the valence band maximum to the three most strongly bound defect states. The calculated energies of the low-energy peaks are similar to those reported in the experimental work of Greben and coworkers who also study MoS$_2$ on an SiO$_2$ substrate~\cite{greben2020} find the first neutral exciton peak at 1.96 eV and the defect-bound exciton peak at 1.77 eV.

\section{Conclusions}
We have calculated the optical absorption spectrum of monolayer MoS$_2$ in the presence of a charged defect by solving the Bethe-Salpeter equation. We find that the presence of the defect gives rise to additional peaks in the spectrum approximately 100 - 200 meV below the A exciton peak in good agreement with experimental observations. These peaks arise from transitions involving bound defect states.   

\section{Acknowledgements} 
This work was supported through a studentship in the Centre for Doctoral Training on Theory and Simulation of Materials at Imperial College London funded by the EPSRC (EP/L015579/1). We acknowledge the Thomas Young Centre under Grant No. TYC-101. 

\subsection{Appendix}
\begin{figure}[ht]
    \centering
    \includegraphics[width=0.49\textwidth]{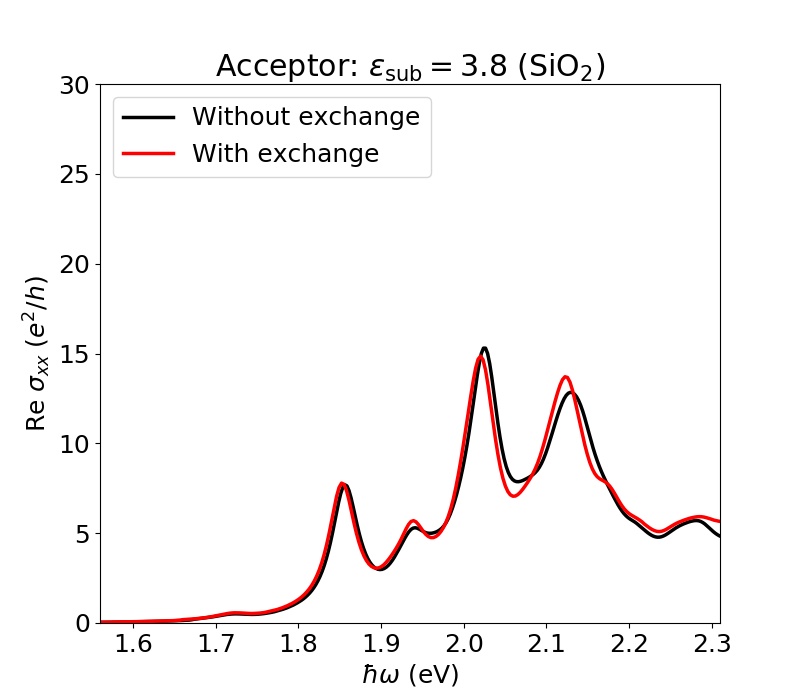}
    \caption{Optical conductivity of an acceptor defect on monolayer MoS$_2$ with (black line) and without (red line) the exchange term of the Bethe-Salpeter equation.}
    \label{fig:optical_exchange}
\end{figure}

Exchange interactions: It is well known that the exchange term of the BSE kernel does not strongly influence the absorption spectrum of the pristine monolayer~\cite{Wu2015}. To test whether this is still the case in the presence of a charged defect, we have calculated the optical conductivity with and without the exchange term for a $12 \times 12$ supercell containing a single acceptor defect, see Projections: Fig.~\ref{fig:optical_exchange}. It is clear that also in the presence of the charged defect, exchange interactions influence the optical spectrum only weakly.

\begin{figure}
    \centering
    \includegraphics[width=0.5\textwidth]{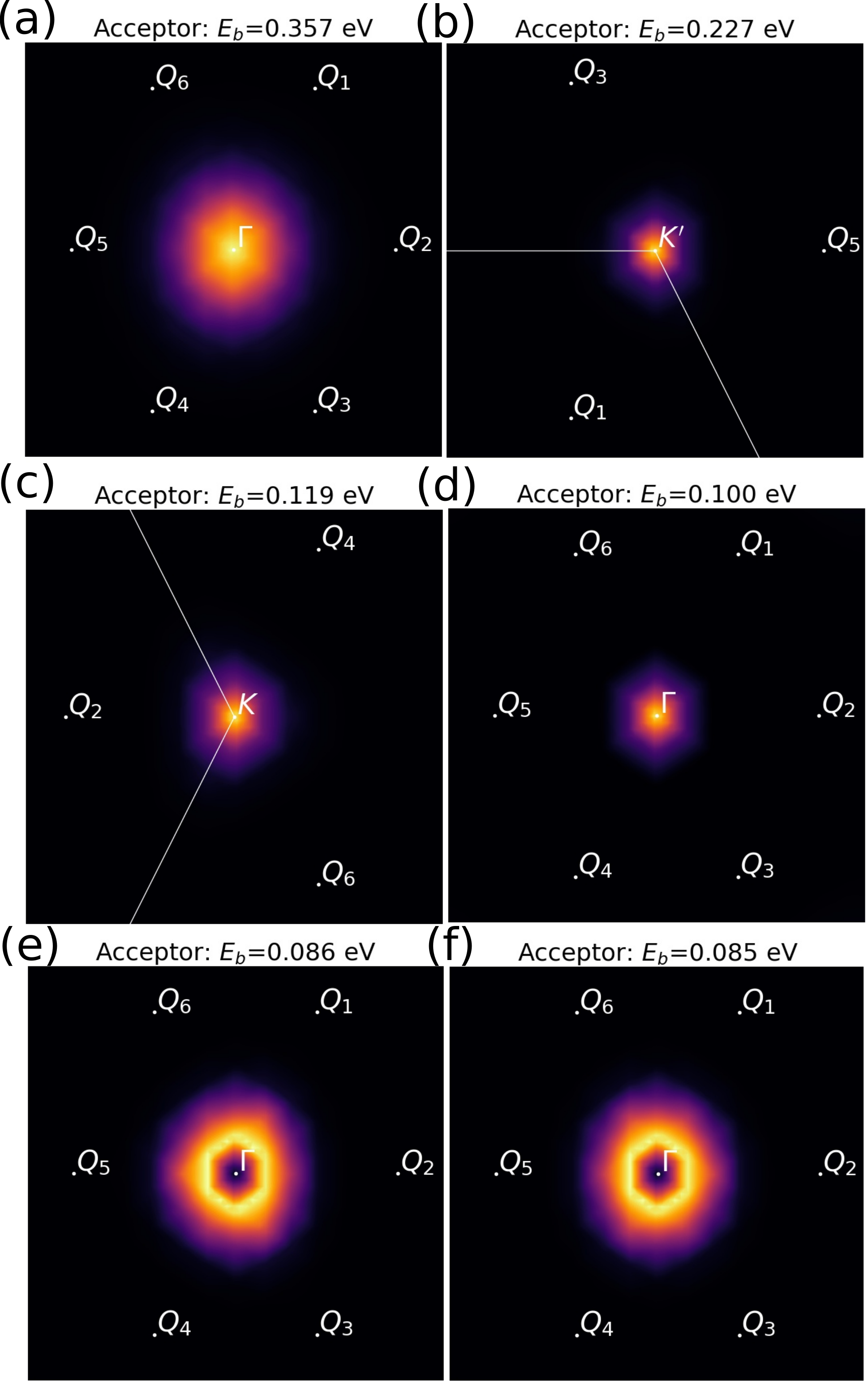}
    \caption{Projections of the most strongly bound acceptor states onto the wavefunctions of the defect-free system.}
    \label{fig:acc_proj}
\end{figure}

\begin{figure}
    \centering
    \includegraphics[width=0.5\textwidth]{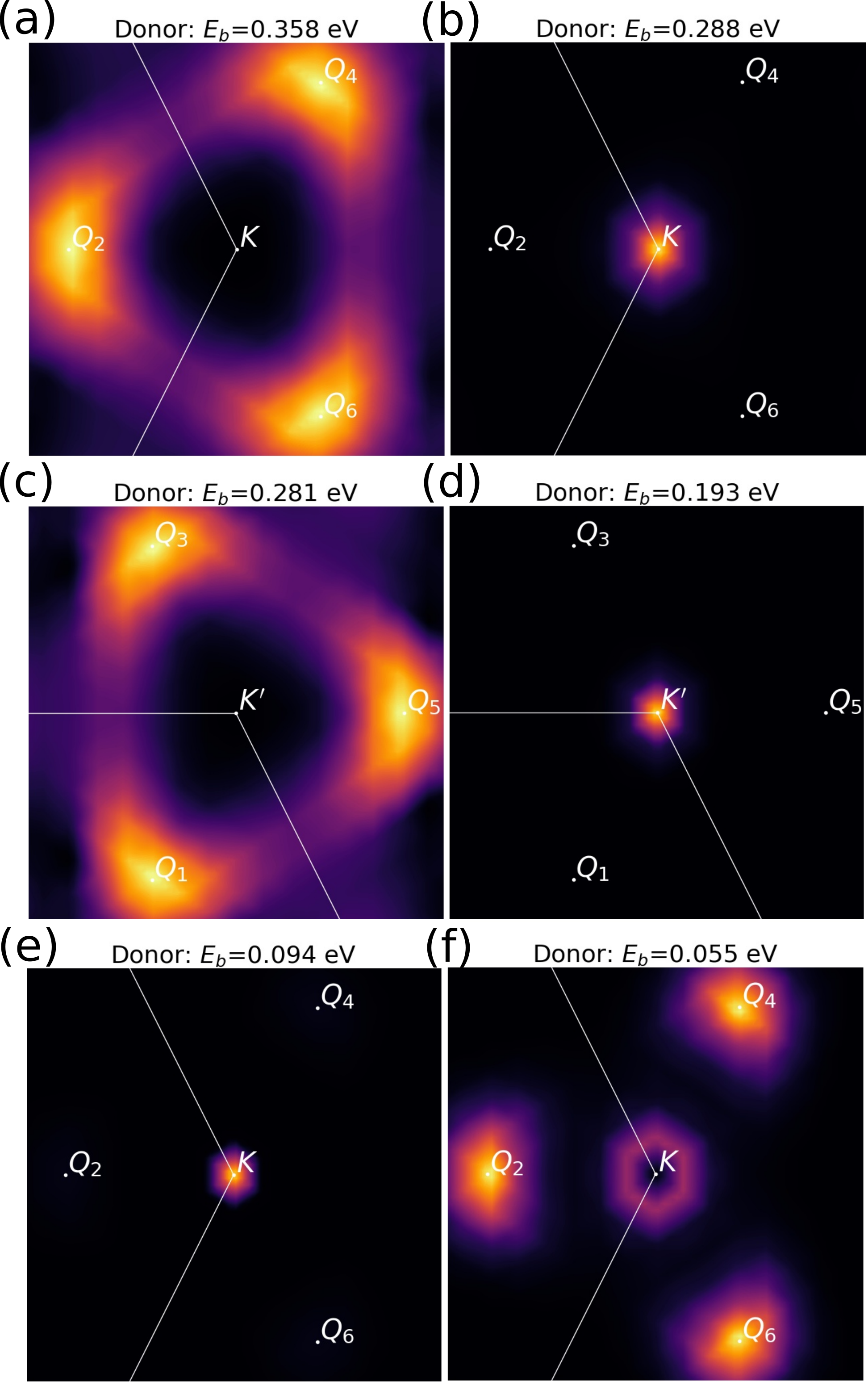}
    \caption{Projections of the most strongly bound donor states onto the wavefunctions of the defect-free system.}
    \label{fig:don_proj}
\end{figure}

Defect state projections: Figures~\ref{fig:acc_proj} and \ref{fig:don_proj} show the projections of acceptor and donor defect states onto the states of the defect-free system, respectively.

Optical matrix elements: Using the tight-binding basis convention
\begin{equation*}
    \psi_{n\mathbf{k}}(\mathbf{x})  = \frac{1}{\sqrt{N_k}}\sum_{li}\tilde{c}_{n\mathbf{k}lj}\sum_{\mathbf{R}}e^{i\mathbf{k}\cdot\mathbf{R}}\phi_{li}^{\mathbf{R}}(\mathbf{r}-\mathbf{R}-\bm{\tau}_j,\alpha),
\end{equation*}
Pedersen \emph{et al.} \cite{Pedersen2001} write the momentum matrix element as
\begin{align*}
    \mathbf{p}_{cv\mathbf{k}} = &\frac{m_0}{\hbar}\sum_{limj}\tilde{c}^*_{c\mathbf{k}li}\tilde{c}_{v\mathbf{k}mj}\nabla_{\mathbf{k}}\tilde{H}_{limj}(\mathbf{k}) \\
    & + \frac{im_0}{\hbar}\left(E_{c\mathbf{k}} - E_{v\mathbf{k}}\right)\sum_{limj}\tilde{c}^*_{c\mathbf{k}li}\tilde{c}_{v\mathbf{k}mj}\mathbf{d}_{limj},
\end{align*}
where $\mathbf{d}_{limj}=\delta_{lm}\delta_{ij}\bm{\tau}_i$ denotes the intra-atomic contribution to the matrix element. In this work, we use a different tight-binding basis convention that includes an additional phase factors $\exp(i \mathbf{k} \cdot \bm{\tau}_j)$, see Eq.~\eqref{eq:basis}. The Hamiltonians and the eigenvectors of the two different conventions are related through~\cite{pythtb}
\begin{align*}
    \tilde{c}_{n\mathbf{k}li} & = c_{n\mathbf{k}li}e^{i\mathbf{k}\cdot \bm{\tau}_i}\\
    \tilde{H}_{limj}(\mathbf{k}) & = H_{limj}(\mathbf{k}) e^{i\mathbf{k}\cdot\left(\bm{\tau}_i-\bm{\tau}_j\right)}.
\end{align*}
Applying this transformation to the expression of the momentum matrix element, we find that
\begin{align*}
    \mathbf{p}_{cv\mathbf{k}} = &\frac{m_0}{\hbar}\sum_{limj}c^*_{c\mathbf{k}li}c_{v\mathbf{k}mj}\left[\nabla_{\mathbf{k}}H_{limj}(\mathbf{k}) +i\left(\bm{\tau}_i-\bm{\tau}_j\right)H_{limj}(\mathbf{k})\right] \\
    & + \frac{im_0}{\hbar}\left(E_{c\mathbf{k}} - E_{v\mathbf{k}}\right)\sum_{li}c^*_{c\mathbf{k}li}c_{v\mathbf{k}li} \bm{\tau}_i \\
    = &\frac{m_0}{\hbar}\sum_{limj}c^*_{c\mathbf{k}li}c_{v\mathbf{k}mj}\nabla_{\mathbf{k}}H_{limj}(\mathbf{k}) \\
    &+\frac{im_0}{\hbar}\sum_{li}c^*_{c\mathbf{k}li}\bm{\tau_i}\left(\sum_{mj}H_{limj}(\mathbf{k})c_{v\mathbf{k}mj}\right)\\
    & - \frac{im_0}{\hbar}\sum_{mj}c_{v\mathbf{k}mj}\bm{\tau}_j\left(\sum_{li} c^*_{c\mathbf{k}li}H_{limj}(\mathbf{k})\right) \\
    & + \frac{im_0}{\hbar}\left(E_{c\mathbf{k}} - E_{v\mathbf{k}}\right)\sum_{li}c^*_{c\mathbf{k}li}c_{v\mathbf{k}li} \bm{\tau}_i\\
    = & \frac{m_0}{\hbar}\sum_{limj}c^*_{c\mathbf{k}li}c_{v\mathbf{k}mj}\nabla_{\mathbf{k}}H_{limj}(\mathbf{k}), 
\end{align*}
i.e. the terms involving $\bm{\tau}_j$ cancel out.

\bibliography{optical}

\begin{thebibliography}{24}%
\makeatletter
\providecommand \@ifxundefined [1]{%
 \@ifx{#1\undefined}
}%
\providecommand \@ifnum [1]{%
 \ifnum #1\expandafter \@firstoftwo
 \else \expandafter \@secondoftwo
 \fi
}%
\providecommand \@ifx [1]{%
 \ifx #1\expandafter \@firstoftwo
 \else \expandafter \@secondoftwo
 \fi
}%
\providecommand \natexlab [1]{#1}%
\providecommand \enquote  [1]{``#1''}%
\providecommand \bibnamefont  [1]{#1}%
\providecommand \bibfnamefont [1]{#1}%
\providecommand \citenamefont [1]{#1}%
\providecommand \href@noop [0]{\@secondoftwo}%
\providecommand \href [0]{\begingroup \@sanitize@url \@href}%
\providecommand \@href[1]{\@@startlink{#1}\@@href}%
\providecommand \@@href[1]{\endgroup#1\@@endlink}%
\providecommand \@sanitize@url [0]{\catcode `\\12\catcode `\$12\catcode
  `\&12\catcode `\#12\catcode `\^12\catcode `\_12\catcode `\%12\relax}%
\providecommand \@@startlink[1]{}%
\providecommand \@@endlink[0]{}%
\providecommand \url  [0]{\begingroup\@sanitize@url \@url }%
\providecommand \@url [1]{\endgroup\@href {#1}{\urlprefix }}%
\providecommand \urlprefix  [0]{URL }%
\providecommand \Eprint [0]{\href }%
\providecommand \doibase [0]{https://doi.org/}%
\providecommand \selectlanguage [0]{\@gobble}%
\providecommand \bibinfo  [0]{\@secondoftwo}%
\providecommand \bibfield  [0]{\@secondoftwo}%
\providecommand \translation [1]{[#1]}%
\providecommand \BibitemOpen [0]{}%
\providecommand \bibitemStop [0]{}%
\providecommand \bibitemNoStop [0]{.\EOS\space}%
\providecommand \EOS [0]{\spacefactor3000\relax}%
\providecommand \BibitemShut  [1]{\csname bibitem#1\endcsname}%
\let\auto@bib@innerbib\@empty
\bibitem [{\citenamefont {Wachter}\ \emph {et~al.}(2017)\citenamefont
  {Wachter}, \citenamefont {Polyushkin}, \citenamefont {Bethge},\ and\
  \citenamefont {Mueller}}]{Wachter2017}%
  \BibitemOpen
  \bibfield  {author} {\bibinfo {author} {\bibfnamefont {S.}~\bibnamefont
  {Wachter}}, \bibinfo {author} {\bibfnamefont {D.~K.}\ \bibnamefont
  {Polyushkin}}, \bibinfo {author} {\bibfnamefont {O.}~\bibnamefont {Bethge}},\
  and\ \bibinfo {author} {\bibfnamefont {T.}~\bibnamefont {Mueller}},\
  }\bibfield  {title} {\bibinfo {title} {A microprocessor based on a
  two-dimensional semiconductor},\ }\href {https://doi.org/10.1038/ncomms14948}
  {\bibfield  {journal} {\bibinfo  {journal} {Nature Communications}\ }\textbf
  {\bibinfo {volume} {8}},\ \bibinfo {pages} {14948} (\bibinfo {year}
  {2017})}\BibitemShut {NoStop}%
\bibitem [{\citenamefont {Lopez-Sanchez}\ \emph {et~al.}(2013)\citenamefont
  {Lopez-Sanchez}, \citenamefont {Lembke}, \citenamefont {Kayci}, \citenamefont
  {Radenovic},\ and\ \citenamefont {Kis}}]{LopezSanchez2013}%
  \BibitemOpen
  \bibfield  {author} {\bibinfo {author} {\bibfnamefont {O.}~\bibnamefont
  {Lopez-Sanchez}}, \bibinfo {author} {\bibfnamefont {D.}~\bibnamefont
  {Lembke}}, \bibinfo {author} {\bibfnamefont {M.}~\bibnamefont {Kayci}},
  \bibinfo {author} {\bibfnamefont {A.}~\bibnamefont {Radenovic}},\ and\
  \bibinfo {author} {\bibfnamefont {A.}~\bibnamefont {Kis}},\ }\bibfield
  {title} {\bibinfo {title} {{Ultrasensitive photodetectors based on monolayer
  MoS$_2$}},\ }\href {https://doi.org/10.1038/nnano.2013.100} {\bibfield
  {journal} {\bibinfo  {journal} {Nature Nanotechnology}\ }\textbf {\bibinfo
  {volume} {8}},\ \bibinfo {pages} {497} (\bibinfo {year} {2013})}\BibitemShut
  {NoStop}%
\bibitem [{\citenamefont {Baugher}\ \emph {et~al.}(2014)\citenamefont
  {Baugher}, \citenamefont {Churchill}, \citenamefont {Yang},\ and\
  \citenamefont {Jarillo-Herrero}}]{Baugher2014}%
  \BibitemOpen
  \bibfield  {author} {\bibinfo {author} {\bibfnamefont {B.~W.~H.}\
  \bibnamefont {Baugher}}, \bibinfo {author} {\bibfnamefont {H.~O.~H.}\
  \bibnamefont {Churchill}}, \bibinfo {author} {\bibfnamefont {Y.}~\bibnamefont
  {Yang}},\ and\ \bibinfo {author} {\bibfnamefont {P.}~\bibnamefont
  {Jarillo-Herrero}},\ }\bibfield  {title} {\bibinfo {title} {Optoelectronic
  devices based on electrically tunable p–n diodes in a monolayer
  dichalcogenide},\ }\href {https://doi.org/10.1038/nnano.2014.25} {\bibfield
  {journal} {\bibinfo  {journal} {Nat. Nano.}\ }\textbf {\bibinfo {volume}
  {9}},\ \bibinfo {pages} {262} (\bibinfo {year} {2014})}\BibitemShut {NoStop}%
\bibitem [{\citenamefont {Liu}\ \emph {et~al.}(2015)\citenamefont {Liu},
  \citenamefont {Xiao}, \citenamefont {Yao}, \citenamefont {Xu},\ and\
  \citenamefont {Yao}}]{liu15}%
  \BibitemOpen
  \bibfield  {author} {\bibinfo {author} {\bibfnamefont {G.-B.}\ \bibnamefont
  {Liu}}, \bibinfo {author} {\bibfnamefont {D.}~\bibnamefont {Xiao}}, \bibinfo
  {author} {\bibfnamefont {Y.}~\bibnamefont {Yao}}, \bibinfo {author}
  {\bibfnamefont {X.}~\bibnamefont {Xu}},\ and\ \bibinfo {author}
  {\bibfnamefont {W.}~\bibnamefont {Yao}},\ }\bibfield  {title} {\bibinfo
  {title} {Electronic structures and theoretical modelling of two-dimensional
  group-vib transition metal dichalcogenides},\ }\href
  {https://doi.org/10.1039/C4CS00301B} {\bibfield  {journal} {\bibinfo
  {journal} {Chem. Soc. Rev.}\ }\textbf {\bibinfo {volume} {44}},\ \bibinfo
  {pages} {2643} (\bibinfo {year} {2015})}\BibitemShut {NoStop}%
\bibitem [{\citenamefont {Kuc}\ \emph {et~al.}(2015)\citenamefont {Kuc},
  \citenamefont {Heine},\ and\ \citenamefont {Kis}}]{kuc15}%
  \BibitemOpen
  \bibfield  {author} {\bibinfo {author} {\bibfnamefont {A.}~\bibnamefont
  {Kuc}}, \bibinfo {author} {\bibfnamefont {T.}~\bibnamefont {Heine}},\ and\
  \bibinfo {author} {\bibfnamefont {A.}~\bibnamefont {Kis}},\ }\bibfield
  {title} {\bibinfo {title} {Electronic properties of transition-metal
  dichalcogenides},\ }\href {https://doi.org/10.1557/mrs.2015.143} {\bibfield
  {journal} {\bibinfo  {journal} {MRS Bulletin}\ }\textbf {\bibinfo {volume}
  {40}},\ \bibinfo {pages} {577} (\bibinfo {year} {2015})}\BibitemShut
  {NoStop}%
\bibitem [{\citenamefont {Mak}\ \emph {et~al.}(2010)\citenamefont {Mak},
  \citenamefont {Lee}, \citenamefont {Hone}, \citenamefont {Shan},\ and\
  \citenamefont {Heinz}}]{Mak2010}%
  \BibitemOpen
  \bibfield  {author} {\bibinfo {author} {\bibfnamefont {K.~F.}\ \bibnamefont
  {Mak}}, \bibinfo {author} {\bibfnamefont {C.}~\bibnamefont {Lee}}, \bibinfo
  {author} {\bibfnamefont {J.}~\bibnamefont {Hone}}, \bibinfo {author}
  {\bibfnamefont {J.}~\bibnamefont {Shan}},\ and\ \bibinfo {author}
  {\bibfnamefont {T.~F.}\ \bibnamefont {Heinz}},\ }\bibfield  {title} {\bibinfo
  {title} {{Atomically thin MoS$_2$: A new direct-gap semiconductor}},\ }\href
  {https://doi.org/10.1103/PhysRevLett.105.136805} {\bibfield  {journal}
  {\bibinfo  {journal} {Phys. Rev. Lett.}\ }\textbf {\bibinfo {volume} {105}},\
  \bibinfo {pages} {136805} (\bibinfo {year} {2010})},\ \Eprint
  {https://arxiv.org/abs/1004.0546} {1004.0546} \BibitemShut {NoStop}%
\bibitem [{\citenamefont {Cao}\ \emph {et~al.}(2012)\citenamefont {Cao},
  \citenamefont {Wang}, \citenamefont {Han}, \citenamefont {Ye}, \citenamefont
  {Zhu}, \citenamefont {Shi}, \citenamefont {Niu}, \citenamefont {Tan},
  \citenamefont {Wang}, \citenamefont {Liu},\ and\ \citenamefont
  {Feng}}]{Cao2012}%
  \BibitemOpen
  \bibfield  {author} {\bibinfo {author} {\bibfnamefont {T.}~\bibnamefont
  {Cao}}, \bibinfo {author} {\bibfnamefont {G.}~\bibnamefont {Wang}}, \bibinfo
  {author} {\bibfnamefont {W.}~\bibnamefont {Han}}, \bibinfo {author}
  {\bibfnamefont {H.}~\bibnamefont {Ye}}, \bibinfo {author} {\bibfnamefont
  {C.}~\bibnamefont {Zhu}}, \bibinfo {author} {\bibfnamefont {J.}~\bibnamefont
  {Shi}}, \bibinfo {author} {\bibfnamefont {Q.}~\bibnamefont {Niu}}, \bibinfo
  {author} {\bibfnamefont {P.}~\bibnamefont {Tan}}, \bibinfo {author}
  {\bibfnamefont {E.}~\bibnamefont {Wang}}, \bibinfo {author} {\bibfnamefont
  {B.}~\bibnamefont {Liu}},\ and\ \bibinfo {author} {\bibfnamefont
  {J.}~\bibnamefont {Feng}},\ }\bibfield  {title} {\bibinfo {title}
  {Valley-selective circular dichroism of monolayer molybdenum disulphide},\
  }\href@noop {} {\bibfield  {journal} {\bibinfo  {journal} {Nature
  Communications}\ }\textbf {\bibinfo {volume} {3}},\ \bibinfo {pages} {887}
  (\bibinfo {year} {2012})}\BibitemShut {NoStop}%
\bibitem [{\citenamefont {Wang}\ \emph {et~al.}(2018)\citenamefont {Wang},
  \citenamefont {Chernikov}, \citenamefont {Glazov}, \citenamefont {Heinz},
  \citenamefont {Marie}, \citenamefont {Amand},\ and\ \citenamefont
  {Urbaszek}}]{Colloquium2018}%
  \BibitemOpen
  \bibfield  {author} {\bibinfo {author} {\bibfnamefont {G.}~\bibnamefont
  {Wang}}, \bibinfo {author} {\bibfnamefont {A.}~\bibnamefont {Chernikov}},
  \bibinfo {author} {\bibfnamefont {M.~M.}\ \bibnamefont {Glazov}}, \bibinfo
  {author} {\bibfnamefont {T.~F.}\ \bibnamefont {Heinz}}, \bibinfo {author}
  {\bibfnamefont {X.}~\bibnamefont {Marie}}, \bibinfo {author} {\bibfnamefont
  {T.}~\bibnamefont {Amand}},\ and\ \bibinfo {author} {\bibfnamefont
  {B.}~\bibnamefont {Urbaszek}},\ }\bibfield  {title} {\bibinfo {title}
  {{Colloquium: Excitons in atomically thin transition metal
  dichalcogenides}},\ }\bibfield  {journal} {\bibinfo  {journal} {Reviews of
  Modern Physics}\ }\textbf {\bibinfo {volume} {90}},\ \href
  {https://doi.org/10.1103/RevModPhys.90.021001} {10.1103/RevModPhys.90.021001}
  (\bibinfo {year} {2018}),\ \Eprint {https://arxiv.org/abs/1707.05863}
  {arXiv:1707.05863} \BibitemShut {NoStop}%
\bibitem [{\citenamefont {Greben}\ \emph {et~al.}(2020)\citenamefont {Greben},
  \citenamefont {Arora}, \citenamefont {Harats},\ and\ \citenamefont
  {Bolotin}}]{greben2020}%
  \BibitemOpen
  \bibfield  {author} {\bibinfo {author} {\bibfnamefont {K.}~\bibnamefont
  {Greben}}, \bibinfo {author} {\bibfnamefont {S.}~\bibnamefont {Arora}},
  \bibinfo {author} {\bibfnamefont {M.~G.}\ \bibnamefont {Harats}},\ and\
  \bibinfo {author} {\bibfnamefont {K.~I.}\ \bibnamefont {Bolotin}},\
  }\bibfield  {title} {\bibinfo {title} {Intrinsic and extrinsic defect-related
  excitons in tmdcs},\ }\href@noop {} {\bibfield  {journal} {\bibinfo
  {journal} {Nano Letters}\ }\textbf {\bibinfo {volume} {20}},\ \bibinfo
  {pages} {2544} (\bibinfo {year} {2020})}\BibitemShut {NoStop}%
\bibitem [{\citenamefont {Shang}\ \emph {et~al.}(2017)\citenamefont {Shang},
  \citenamefont {Cong}, \citenamefont {Shen}, \citenamefont {Yang},
  \citenamefont {Zou}, \citenamefont {Peimyoo}, \citenamefont {Cao},
  \citenamefont {Eginligil}, \citenamefont {Lin}, \citenamefont {Huang} \emph
  {et~al.}}]{shang2017}%
  \BibitemOpen
  \bibfield  {author} {\bibinfo {author} {\bibfnamefont {J.}~\bibnamefont
  {Shang}}, \bibinfo {author} {\bibfnamefont {C.}~\bibnamefont {Cong}},
  \bibinfo {author} {\bibfnamefont {X.}~\bibnamefont {Shen}}, \bibinfo {author}
  {\bibfnamefont {W.}~\bibnamefont {Yang}}, \bibinfo {author} {\bibfnamefont
  {C.}~\bibnamefont {Zou}}, \bibinfo {author} {\bibfnamefont {N.}~\bibnamefont
  {Peimyoo}}, \bibinfo {author} {\bibfnamefont {B.}~\bibnamefont {Cao}},
  \bibinfo {author} {\bibfnamefont {M.}~\bibnamefont {Eginligil}}, \bibinfo
  {author} {\bibfnamefont {W.}~\bibnamefont {Lin}}, \bibinfo {author}
  {\bibfnamefont {W.}~\bibnamefont {Huang}}, \emph {et~al.},\ }\bibfield
  {title} {\bibinfo {title} {Revealing electronic nature of broad bound exciton
  bands in two-dimensional semiconducting w s 2 and mo s 2},\ }\href@noop {}
  {\bibfield  {journal} {\bibinfo  {journal} {Physical Review Materials}\
  }\textbf {\bibinfo {volume} {1}},\ \bibinfo {pages} {074001} (\bibinfo {year}
  {2017})}\BibitemShut {NoStop}%
\bibitem [{\citenamefont {Ganchev}\ \emph {et~al.}(2015)\citenamefont
  {Ganchev}, \citenamefont {Drummond}, \citenamefont {Aleiner},\ and\
  \citenamefont {Fal’Ko}}]{ganchev2015}%
  \BibitemOpen
  \bibfield  {author} {\bibinfo {author} {\bibfnamefont {B.}~\bibnamefont
  {Ganchev}}, \bibinfo {author} {\bibfnamefont {N.}~\bibnamefont {Drummond}},
  \bibinfo {author} {\bibfnamefont {I.}~\bibnamefont {Aleiner}},\ and\ \bibinfo
  {author} {\bibfnamefont {V.}~\bibnamefont {Fal’Ko}},\ }\bibfield  {title}
  {\bibinfo {title} {Three-particle complexes in two-dimensional
  semiconductors},\ }\href@noop {} {\bibfield  {journal} {\bibinfo  {journal}
  {Physical review letters}\ }\textbf {\bibinfo {volume} {114}},\ \bibinfo
  {pages} {107401} (\bibinfo {year} {2015})}\BibitemShut {NoStop}%
\bibitem [{\citenamefont {Wu}(2022)}]{wu2022}%
  \BibitemOpen
  \bibfield  {author} {\bibinfo {author} {\bibfnamefont {S.-D.}\ \bibnamefont
  {Wu}},\ }\bibfield  {title} {\bibinfo {title} {Hydrogenic donor impurity
  states and intersubband optical absorption spectra in monolayer transition
  metal dichalcogenides with dielectric environments},\ }\href@noop {}
  {\bibfield  {journal} {\bibinfo  {journal} {Chinese Physics B}\ } (\bibinfo
  {year} {2022})}\BibitemShut {NoStop}%
\bibitem [{\citenamefont {Aghajanian}\ \emph {et~al.}(2018)\citenamefont
  {Aghajanian}, \citenamefont {Mostofi},\ and\ \citenamefont
  {Lischner}}]{Aghajanian2018}%
  \BibitemOpen
  \bibfield  {author} {\bibinfo {author} {\bibfnamefont {M.}~\bibnamefont
  {Aghajanian}}, \bibinfo {author} {\bibfnamefont {A.~A.}\ \bibnamefont
  {Mostofi}},\ and\ \bibinfo {author} {\bibfnamefont {J.}~\bibnamefont
  {Lischner}},\ }\bibfield  {title} {\bibinfo {title} {{Tuning electronic
  properties of transition-metal dichalcogenides via defect charge}},\ }\href
  {https://doi.org/10.1038/s41598-018-31941-1} {\bibfield  {journal} {\bibinfo
  {journal} {Scientific Reports}\ }\textbf {\bibinfo {volume} {8}},\ \bibinfo
  {pages} {13611} (\bibinfo {year} {2018})}\BibitemShut {NoStop}%
\bibitem [{\citenamefont {Aghajanian}\ \emph {et~al.}(2020)\citenamefont
  {Aghajanian}, \citenamefont {Schuler}, \citenamefont {Cochrane},
  \citenamefont {Lee}, \citenamefont {Kastl}, \citenamefont {Neaton},
  \citenamefont {Weber-Bargioni}, \citenamefont {Mostofi},\ and\ \citenamefont
  {Lischner}}]{Aghajanian2020}%
  \BibitemOpen
  \bibfield  {author} {\bibinfo {author} {\bibfnamefont {M.}~\bibnamefont
  {Aghajanian}}, \bibinfo {author} {\bibfnamefont {B.}~\bibnamefont {Schuler}},
  \bibinfo {author} {\bibfnamefont {K.~A.}\ \bibnamefont {Cochrane}}, \bibinfo
  {author} {\bibfnamefont {J.-H.}\ \bibnamefont {Lee}}, \bibinfo {author}
  {\bibfnamefont {C.}~\bibnamefont {Kastl}}, \bibinfo {author} {\bibfnamefont
  {J.~B.}\ \bibnamefont {Neaton}}, \bibinfo {author} {\bibfnamefont
  {A.}~\bibnamefont {Weber-Bargioni}}, \bibinfo {author} {\bibfnamefont
  {A.~A.}\ \bibnamefont {Mostofi}},\ and\ \bibinfo {author} {\bibfnamefont
  {J.}~\bibnamefont {Lischner}},\ }\bibfield  {title} {\bibinfo {title}
  {{Resonant and bound states of charged defects in two-dimensional
  semiconductors}},\ }\href {https://doi.org/10.1103/physrevb.101.081201}
  {\bibfield  {journal} {\bibinfo  {journal} {Phys. Rev. B}\ }\textbf {\bibinfo
  {volume} {101}},\ \bibinfo {pages} {081201} (\bibinfo {year}
  {2020})}\BibitemShut {NoStop}%
\bibitem [{\citenamefont {Poellmann}\ \emph {et~al.}(2015)\citenamefont
  {Poellmann}, \citenamefont {Steinleitner}, \citenamefont {Leierseder},
  \citenamefont {Nagler}, \citenamefont {Plechinger}, \citenamefont {Porer},
  \citenamefont {Bratschitsch}, \citenamefont {Sch{\"u}ller}, \citenamefont
  {Korn},\ and\ \citenamefont {Huber}}]{Poellmann2015}%
  \BibitemOpen
  \bibfield  {author} {\bibinfo {author} {\bibfnamefont {C.}~\bibnamefont
  {Poellmann}}, \bibinfo {author} {\bibfnamefont {P.}~\bibnamefont
  {Steinleitner}}, \bibinfo {author} {\bibfnamefont {U.}~\bibnamefont
  {Leierseder}}, \bibinfo {author} {\bibfnamefont {P.}~\bibnamefont {Nagler}},
  \bibinfo {author} {\bibfnamefont {G.}~\bibnamefont {Plechinger}}, \bibinfo
  {author} {\bibfnamefont {M.}~\bibnamefont {Porer}}, \bibinfo {author}
  {\bibfnamefont {R.}~\bibnamefont {Bratschitsch}}, \bibinfo {author}
  {\bibfnamefont {C.}~\bibnamefont {Sch{\"u}ller}}, \bibinfo {author}
  {\bibfnamefont {T.}~\bibnamefont {Korn}},\ and\ \bibinfo {author}
  {\bibfnamefont {R.}~\bibnamefont {Huber}},\ }\bibfield  {title} {\bibinfo
  {title} {Resonant internal quantum transitions and femtosecond radiative
  decay of excitons in monolayer wse2},\ }\href@noop {} {\bibfield  {journal}
  {\bibinfo  {journal} {Nature Materials}\ }\textbf {\bibinfo {volume} {14}},\
  \bibinfo {pages} {889} (\bibinfo {year} {2015})}\BibitemShut {NoStop}%
\bibitem [{\citenamefont {Rigosi}\ \emph {et~al.}(2016)\citenamefont {Rigosi},
  \citenamefont {Hill}, \citenamefont {Rim}, \citenamefont {Flynn},\ and\
  \citenamefont {Heinz}}]{Rigosi2016}%
  \BibitemOpen
  \bibfield  {author} {\bibinfo {author} {\bibfnamefont {A.~F.}\ \bibnamefont
  {Rigosi}}, \bibinfo {author} {\bibfnamefont {H.~M.}\ \bibnamefont {Hill}},
  \bibinfo {author} {\bibfnamefont {K.~T.}\ \bibnamefont {Rim}}, \bibinfo
  {author} {\bibfnamefont {G.~W.}\ \bibnamefont {Flynn}},\ and\ \bibinfo
  {author} {\bibfnamefont {T.~F.}\ \bibnamefont {Heinz}},\ }\bibfield  {title}
  {\bibinfo {title} {Electronic band gaps and exciton binding energies in
  monolayer
  $\mathrm{M}{\mathrm{o}}_{x}{\mathrm{w}}_{1\text{\ensuremath{-}}x}{\mathrm{s}}_{2}$
  transition metal dichalcogenide alloys probed by scanning tunneling and
  optical spectroscopy},\ }\href {https://doi.org/10.1103/PhysRevB.94.075440}
  {\bibfield  {journal} {\bibinfo  {journal} {Phys. Rev. B}\ }\textbf {\bibinfo
  {volume} {94}},\ \bibinfo {pages} {075440} (\bibinfo {year}
  {2016})}\BibitemShut {NoStop}%
\bibitem [{\citenamefont {Zhang}\ \emph {et~al.}(2015)\citenamefont {Zhang},
  \citenamefont {Wan}, \citenamefont {Ma}, \citenamefont {Wang}, \citenamefont
  {Wang},\ and\ \citenamefont {Dai}}]{Zhang2015}%
  \BibitemOpen
  \bibfield  {author} {\bibinfo {author} {\bibfnamefont {H.}~\bibnamefont
  {Zhang}}, \bibinfo {author} {\bibfnamefont {Y.}~\bibnamefont {Wan}}, \bibinfo
  {author} {\bibfnamefont {Y.}~\bibnamefont {Ma}}, \bibinfo {author}
  {\bibfnamefont {W.}~\bibnamefont {Wang}}, \bibinfo {author} {\bibfnamefont
  {Y.}~\bibnamefont {Wang}},\ and\ \bibinfo {author} {\bibfnamefont
  {L.}~\bibnamefont {Dai}},\ }\bibfield  {title} {\bibinfo {title}
  {Interference effect on optical signals of monolayer mos2},\ }\href@noop {}
  {\bibfield  {journal} {\bibinfo  {journal} {Applied Physics Letters}\
  }\textbf {\bibinfo {volume} {107}},\ \bibinfo {pages} {101904} (\bibinfo
  {year} {2015})}\BibitemShut {NoStop}%
\bibitem [{\citenamefont {Wu}\ \emph {et~al.}(2015)\citenamefont {Wu},
  \citenamefont {Qu},\ and\ \citenamefont {MacDonald}}]{Wu2015}%
  \BibitemOpen
  \bibfield  {author} {\bibinfo {author} {\bibfnamefont {F.}~\bibnamefont
  {Wu}}, \bibinfo {author} {\bibfnamefont {F.}~\bibnamefont {Qu}},\ and\
  \bibinfo {author} {\bibfnamefont {A.~H.}\ \bibnamefont {MacDonald}},\
  }\bibfield  {title} {\bibinfo {title} {{Exciton band structure of monolayer
  MoS$_2$}},\ }\href {https://doi.org/10.1103/PhysRevB.91.075310} {\bibfield
  {journal} {\bibinfo  {journal} {Phys. Rev. B}\ }\textbf {\bibinfo {volume}
  {91}},\ \bibinfo {pages} {075310} (\bibinfo {year} {2015})}\BibitemShut
  {NoStop}%
\bibitem [{\citenamefont {Ridolfi}\ \emph {et~al.}(2018)\citenamefont
  {Ridolfi}, \citenamefont {Lewenkopf},\ and\ \citenamefont
  {Pereira}}]{Ridolfi2018}%
  \BibitemOpen
  \bibfield  {author} {\bibinfo {author} {\bibfnamefont {E.}~\bibnamefont
  {Ridolfi}}, \bibinfo {author} {\bibfnamefont {C.~H.}\ \bibnamefont
  {Lewenkopf}},\ and\ \bibinfo {author} {\bibfnamefont {V.~M.}\ \bibnamefont
  {Pereira}},\ }\bibfield  {title} {\bibinfo {title} {{Excitonic structure of
  the optical conductivity in MoS$_2$ monolayers}},\ }\href
  {https://doi.org/10.1103/PhysRevB.97.205409} {\bibfield  {journal} {\bibinfo
  {journal} {Physical Review B}\ }\textbf {\bibinfo {volume} {97}},\ \bibinfo
  {pages} {205409} (\bibinfo {year} {2018})},\ \Eprint
  {https://arxiv.org/abs/1801.07974} {1801.07974} \BibitemShut {NoStop}%
\bibitem [{\citenamefont {Qiu}\ \emph {et~al.}(2016)\citenamefont {Qiu},
  \citenamefont {da~Jornada F.~H.},\ and\ \citenamefont {Louie}}]{Qiu2016}%
  \BibitemOpen
  \bibfield  {author} {\bibinfo {author} {\bibfnamefont {D.~Y.}\ \bibnamefont
  {Qiu}}, \bibinfo {author} {\bibnamefont {da~Jornada F.~H.}},\ and\ \bibinfo
  {author} {\bibfnamefont {S.~G.}\ \bibnamefont {Louie}},\ }\bibfield  {title}
  {\bibinfo {title} {{Screening and many-body effects in two-dimensional
  crystals: Monolayer MoS$_2$}},\ }\href
  {https://doi.org/10.1103/PhysRevB.93.235435} {\bibfield  {journal} {\bibinfo
  {journal} {Phys. Rev. B}\ }\textbf {\bibinfo {volume} {93}},\ \bibinfo
  {pages} {235435} (\bibinfo {year} {2016})}\BibitemShut {NoStop}%
\bibitem [{\citenamefont {Liu}\ \emph {et~al.}(2013)\citenamefont {Liu},
  \citenamefont {Shan}, \citenamefont {Yao}, \citenamefont {Yao},\ and\
  \citenamefont {Xiao}}]{liu13}%
  \BibitemOpen
  \bibfield  {author} {\bibinfo {author} {\bibfnamefont {G.-B.}\ \bibnamefont
  {Liu}}, \bibinfo {author} {\bibfnamefont {W.-Y.}\ \bibnamefont {Shan}},
  \bibinfo {author} {\bibfnamefont {Y.}~\bibnamefont {Yao}}, \bibinfo {author}
  {\bibfnamefont {W.}~\bibnamefont {Yao}},\ and\ \bibinfo {author}
  {\bibfnamefont {D.}~\bibnamefont {Xiao}},\ }\bibfield  {title} {\bibinfo
  {title} {Three-band tight-binding model for monolayers of group-vib
  transition metal dichalcogenides},\ }\href
  {https://doi.org/10.1103/PhysRevB.88.085433} {\bibfield  {journal} {\bibinfo
  {journal} {Phys. Rev. B}\ }\textbf {\bibinfo {volume} {88}},\ \bibinfo
  {pages} {085433} (\bibinfo {year} {2013})}\BibitemShut {NoStop}%
\bibitem [{\citenamefont {Pedersen}\ \emph {et~al.}(2001)\citenamefont
  {Pedersen}, \citenamefont {Pedersen},\ and\ \citenamefont
  {Brun~Kriestensen}}]{Pedersen2001}%
  \BibitemOpen
  \bibfield  {author} {\bibinfo {author} {\bibfnamefont {T.~G.}\ \bibnamefont
  {Pedersen}}, \bibinfo {author} {\bibfnamefont {K.}~\bibnamefont {Pedersen}},\
  and\ \bibinfo {author} {\bibfnamefont {T.}~\bibnamefont {Brun~Kriestensen}},\
  }\bibfield  {title} {\bibinfo {title} {Optical matrix elements in
  tight-binding calculations},\ }\href
  {https://doi.org/10.1103/PhysRevB.63.201101} {\bibfield  {journal} {\bibinfo
  {journal} {Phys. Rev. B}\ }\textbf {\bibinfo {volume} {63}},\ \bibinfo
  {pages} {201101} (\bibinfo {year} {2001})}\BibitemShut {NoStop}%
\bibitem [{\citenamefont {Cruz}\ \emph {et~al.}(1999)\citenamefont {Cruz},
  \citenamefont {Beltr\'an}, \citenamefont {Wang}, \citenamefont {Tag\"ue\~na
  Mart\'{\i}nez},\ and\ \citenamefont {Rubo}}]{Cruz1999}%
  \BibitemOpen
  \bibfield  {author} {\bibinfo {author} {\bibfnamefont {M.}~\bibnamefont
  {Cruz}}, \bibinfo {author} {\bibfnamefont {M.~R.}\ \bibnamefont {Beltr\'an}},
  \bibinfo {author} {\bibfnamefont {C.}~\bibnamefont {Wang}}, \bibinfo {author}
  {\bibfnamefont {J.}~\bibnamefont {Tag\"ue\~na Mart\'{\i}nez}},\ and\ \bibinfo
  {author} {\bibfnamefont {Y.~G.}\ \bibnamefont {Rubo}},\ }\bibfield  {title}
  {\bibinfo {title} {Supercell approach to the optical properties of porous
  silicon},\ }\href {https://doi.org/10.1103/PhysRevB.59.15381} {\bibfield
  {journal} {\bibinfo  {journal} {Phys. Rev. B}\ }\textbf {\bibinfo {volume}
  {59}},\ \bibinfo {pages} {15381} (\bibinfo {year} {1999})}\BibitemShut
  {NoStop}%
\bibitem [{\citenamefont {Yusufaly}\ \emph {et~al.}(2022)\citenamefont
  {Yusufaly}, \citenamefont {Vanderbilt},\ and\ \citenamefont {Coh}}]{pythtb}%
  \BibitemOpen
  \bibfield  {author} {\bibinfo {author} {\bibfnamefont {T.}~\bibnamefont
  {Yusufaly}}, \bibinfo {author} {\bibfnamefont {D.}~\bibnamefont
  {Vanderbilt}},\ and\ \bibinfo {author} {\bibfnamefont {S.}~\bibnamefont
  {Coh}},\ }\bibfield  {title} {\bibinfo {title} {Tight-binding formalism in
  the context of the pythtb package},\ }\href@noop {} {\bibfield  {journal}
  {\bibinfo  {journal} {https://www.physics.rutgers.edu/pythtb/formalism.html}\
  } (\bibinfo {year} {2022})}\BibitemShut {NoStop}%
\end{thebibliography}%
\end{document}